\documentclass[english,preprint,superscriptaddress]{revtex4-1}
\usepackage[T1]{fontenc}
\usepackage[latin9]{inputenc}
\usepackage{color}
\usepackage{graphicx}
\usepackage{esint}

\makeatletter
 
 \@ifundefined{textcolor}{}
 {%
   \definecolor{BLACK}{gray}{0}
   \definecolor{WHITE}{gray}{1}
   \definecolor{RED}{rgb}{1,0,0}
   \definecolor{GREEN}{rgb}{0,1,0}
   \definecolor{BLUE}{rgb}{0,0,1}
   \definecolor{CYAN}{cmyk}{1,0,0,0}
   \definecolor{MAGENTA}{cmyk}{0,1,0,0}
   \definecolor{YELLOW}{cmyk}{0,0,1,0}
 }

 \usepackage[english]{babel}

\makeatother

\usepackage{babel}
\begin{document}

\title{Plasmoid Instability in High-Lundquist-Number Magnetic Reconnection}

\author{Yi-Min Huang }

\email{yimin.huang@unh.edu}

\selectlanguage{english}%

\affiliation{Center for Integrated Computation and Analysis of Reconnection and
Turbulence }

\affiliation{Center for Magnetic Self-Organization in Laboratory and Astrophysical
Plasmas}

\affiliation{Space Science Center, University of New Hampshire, Durham, NH 03824}

\author{A. Bhattacharjee}

\affiliation{Center for Integrated Computation and Analysis of Reconnection and
Turbulence }

\affiliation{Center for Magnetic Self-Organization in Laboratory and Astrophysical
Plasmas}

\affiliation{Max Planck-Princeton Research Center for Plasma Physics and Princeton
Plasma Physics Laboratory, Princeton, NJ 08543}
\begin{abstract}
Our understanding of magnetic reconnection in resistive magnetohydrodynamics
has gone through a fundamental change in recent years. The conventional
wisdom is that magnetic reconnection mediated by resistivity is slow
in laminar high Lundquist ($S$) plasmas, constrained by the scaling
of the reconnection rate predicted by Sweet-Parker theory. However,
recent studies have shown that when $S$ exceeds a critical value
$\sim10^{4}$, the Sweet-Parker current sheet is unstable to a super-Alfv\'enic
plasmoid instability, with a{{} linear} growth rate
that scales as $S^{1/4}$. In the fully developed statistical steady
state {of two-dimensional resistive magnetohydrodynamic
simulations, the normalized average reconnection rate is approximately
$0.01$,} nearly independent of $S$, and the distribution function
$f(\psi)$ of plasmoid magnetic flux $\psi$ follows a power law $f(\psi)\sim\psi^{-1}$.
When Hall effects are included, the plasmoid instability may trigger
onset of Hall reconnection even when the conventional criterion for
onset is not satisfied. The rich variety of possible reconnection
dynamics  is organized in the framework of a phase diagram. 
\end{abstract}
\maketitle

\section{Introduction}

Magnetic reconnection is generally believed to be the underlying mechanism
that powers explosive events such as flares, substorms, and sawtooth
crashes in fusion plasmas.\cite{ZweibelY2009,YamadaKJ2010} Traditionally,
magnetic reconnection is grossly classified into two categories, i.e.
collisional and collisionless reconnection, depending on whether ions
and electrons remain coupled or not in the diffusion region. The conventional
wisdom is that laminar collisional reconnection is described by the
classical Sweet-Parker theory,\cite{Sweet1958,Parker1957} which assumes
an elongated current sheet characterized by a length $L$ of the order
of the system size. The governing dimensionless parameter is the Lundquist
number $S\equiv V_{A}L/\eta$, where $V_{A}$ is the upstream Alfv\'en
speed, $L$ is the reconnection layer length, and $\eta$ is the resistivity.
According to Sweet-Parker theory, the reconnection rate $\sim BV_{A}/\sqrt{S}$,
where $B$ is the upstream magnetic field. In many systems of interest,
the Lundquist number $S$ is very high (e.g. $S\sim10^{12}-10^{14}$
in solar corona), and the corresponding Sweet-Parker reconnection
rate is too slow to account for energy release events. For this reason,
research on fast reconnection in the past two decades has mostly focused
on collisionless reconnection, which can yield reconnection rates
as fast as $\sim0.1V_{A}B$.\cite{BirnDSRDHKMBOP2001} 

The Sweet-Parker theory assumes that the elongated current sheet is
stable. However, it has long been known that this assumption may not
hold, because the current sheet can become unstable to tearing instability
and spontaneously form plasmoids at high-$S${.\cite{Jaggi1963,BulanovSS1979,LeeF1986,Biskamp1986,YanLP1992,ShibataT2001,Lapenta2008}}
Biskamp estimated the critical Lundquist number $S_{c}$ to be approximately
$10^{4}$.{\cite{Biskamp1986}} {Observational
evidences of plasmoids have been reported in the Earth's magnetosphere
and the solar atmosphere}.\cite{OhyamaS1998,KliemKB2000,KoRLLLF2003,AsaiYSS2004,SuiHD2005,LinCF2008,NishizukaTAS2010,Chen_et_al2008,ChenBPYBIMDLKVFG2008,LiuJWLY2010,BartaBKK2011}
Although numerical and observational evidences of plasmoids were abundant,
precise scalings of of the linear instability (hereafter the plasmoid
instability) were not known until recently. In a seminal paper, Loureiro
\emph{et al. }predicted that the linear growth rate $\gamma$ scales
as $\gamma\sim S^{1/4}L/V_{A}$, and the number of plasmoids scales
as $S^{3/8}$.\cite{LoureiroSC2007} At first sight, these results
are rather counterintuitive, because linear growth rates of most resistive
instabilities scale with $S$ to some negative fractional power indices
instead of positive ones. The crucial point is to realize that the
equilibrium, i.e. the Sweet-Parker current sheet, also scales with
$S$.{{} The results of Loureiro }{\emph{et
al.}}{{} can be readily derived once the Sweet-Parker
scaling of the current sheet width $\delta_{SP}\sim L/\sqrt{S}$ is
incorporated in the classical tearing mode dispersion relation.\cite{BhattacharjeeHYR2009} }

While the work of Loureiro \emph{et al.} drew attention to the surprising
scaling properties of the linear plasmoid instability,\cite{LoureiroSC2007}
it is the nonlinear behavior of the instability that has proved to
be transformational for traditional reconnection theory. The onset
of the linear plasmoid instability would not be nearly as interesting
were it not for the fact that it leads to a new nonlinear regime,\cite{BhattacharjeeHYR2009,LoureiroUSCY2009,CassakSD2009,HuangB2010,UzdenskyLS2010}
entirely unanticipated by Sweet-Parker reconnection theory. Furthermore,
the plasmoid instability, which appears to be ubiquitous in high-$S$
plasmas, often leads inevitably to kinetic regimes in which the onset
of fast reconnection occurs earlier than was previously thought possible.\cite{DaughtonRAKYB2009,ShepherdC2010,HuangBS2011}
The picture emerging from these recent studies is that large-scale,
high-Lundquist-number magnetic reconnection can exhibit complex and
rich dynamics that we are only beginning to understand.

The main objective of this paper is to give our perspective of the
recent advances in this subject, and address some issues that have
arisen during the course of this work but have not been addressed
in previous publications. {Most of our discussion
is limited to two-dimensional (2D) systems. The plasmoid instability
in three-dimensional (3D) systems, which is currently an active area
of research, will be briefly discussed.} This paper is organized as
follows. Section \ref{sec:Linear-Theory} reviews the linear theory,
with emphasis on the convective nature of the instability that is
one of its key features. Section \ref{sec:Nonlinear} discusses plasmoid
dynamics in the fully nonlinear regime. Results on reconnection rate
and scaling laws are reviewed, and a heuristic argument is given.
Section \ref{sec:Statistical-Distribution} discusses statistical
descriptions of the plasmoid distribution, which has been a topic
of considerable interest and debate in recent years. We approach this
problem with a combination of analytical models of plasmoid kinetics
and direct numerical simulations (DNS), where the plasmoid distribution
function is found to obey a power law of index $-1$ for smaller plasmoids,
followed by an exponential falloff for large plasmoids. The condition
for transition from the power-law regime to the exponential tail is
discussed in great detail, which is crucial for interpreting the numerical
results. Section \ref{sec:Roles-of-Plasmoid} addresses the role of
the plasmoid instability on the onset of collisionless reconnection.
A revised phase diagram is presented that explicitly includes the
physics of bistability \cite{CassakSD2005} and the intermediate regime
reported in Ref. \cite{HuangBS2011}. Although these effects were
discussed before, they were omitted in our previous rendition of the
phase diagram. Finally, open questions and future challenges are summarized
in Sec. \ref{sec:Discussion-and-Future}.

\section{Linear Theory of the Plasmoid Instability\label{sec:Linear-Theory}}

Consider a Harris sheet of width $a$ with the equilibrium magnetic
field $\mathbf{B}=B_{0}\tanh(z/a)\mathbf{\hat{x}}$. According to
the classical linear tearing instability theory, the tearing mode
growth rate for the \textquotedblleft{}constant-$\psi$\textquotedblright{}
and \textquotedblleft{}non- constant-$\psi$\textquotedblright{} regimes
are as follows, respectively:\cite{CoppiGPRR1976}
\begin{equation}
\gamma\sim\frac{V_{A}}{a}\times\left\{ \begin{array}{l}
S_{a}^{-3/5}\left(ka\right)^{-2/5}\left(1-k^{2}a^{2}\right)^{4/5},\, ka\gg S_{a}^{-1/4}\\
S_{a}^{-1/3}\left(ka\right)^{2/3},\, ka\ll S_{a}^{-1/4}
\end{array}\right.\label{eq:tearing}
\end{equation}
where $S_{a}=aV_{A}/\eta$ is the Lundquist number based on the current
sheet width $a$. The transition from \textquotedblleft{}constant-$\psi$\textquotedblright{}
to \textquotedblleft{}non-constant-$\psi$\textquotedblright{} modes
occurs at $k_{max}a\sim S_{a}^{-1/4}$, where the linear growth rate
peaks and scales as $\gamma_{max}\sim S_{a}^{-1/2}(V_{A}/a)$. Note
that the linear growth rate is proportional to $S_{a}$ raised to
a negative fractional exponent in both regimes. To apply the tearing
mode theory to a Sweet-Parker current sheet, it is important to note
that the current sheet length $L$ is dictated by the global geometry,
and remains approximately the same when $\eta$ varies. On the other
hand, the current sheet width $\delta_{SP}$ follows the scaling $\delta_{SP}\sim L/\sqrt{S}$,
where the Lundquist number $S\equiv LV_{A}/\eta$ is now defined with
the current sheet length. The current sheet width $a$ in should now
be replaced by $\delta_{SP}$, and $S_{a}$ is related to $S$ via
the relation $S_{a}=\delta_{SP}V_{A}/\eta=S^{1/2}$. After some algebra,
the dispersion relation (\ref{eq:tearing}) can be rewritten as 
\begin{equation}
\gamma\sim\frac{V_{A}}{L}\times\left\{ \begin{array}{cc}
S^{2/5}\kappa^{-2/5}(1-\kappa^{2}\epsilon^{2})^{4/5}, & \kappa\gg S^{3/8}\\
\kappa^{2/3}, & \kappa\ll S^{3/8}
\end{array}\right.,\label{eq:plasmoid_dispersion}
\end{equation}
where $\kappa\equiv kL$ and $\epsilon\equiv\delta_{SP}/L\sim S^{-1/2}$.
The peak growth rate occurs at $\kappa_{max}\sim S^{3/8}$, with $\gamma_{max}\sim S^{1/4}(V_{A}/L)$.
The number of plasmoids generated in the linear regime can be estimated
as $n_{p}^{L}\sim L/\lambda_{max}\sim\kappa_{max}\sim S^{3/8}$, where
$\lambda_{max}$ is the wavelength of the fastest growing mode. These
scaling relations have been confirmed by numerical studies.\cite{SamtaneyLUSC2009,NiGHSYB2010,HuangB2010}
The plasmoid instability is referred to as a super-Alfv\'enic instability,\cite{BhattacharjeeHYR2009}
because the linear growth rate far exceeds the inverse of the Alfv\'en
time scale $\tau_{A}\equiv L/V_{A}$ along the current sheet in the
high-$S$ limit. It should be noted, however, that the growth rate
is always slower than the inverse of the Alfv\'en time scale $\delta_{SP}/V_{A}$
transverse to the current sheet, as is required for the tearing mode
analysis. 

The above analysis (also the analysis in Ref. \cite{LoureiroSC2007})
ignores the effects of flow, which is self-consistently generated
in a Sweet-Parker current sheet, on the instability. However, it is
important to appreciate the convective nature of the instability.
Because the outflow speed is of the order of $V_{A}$, disturbance
in the current sheet will be convected out at the Alfv\'en time scale
$\tau_{A}$. Based on the scaling $\gamma_{max}\sim S^{1/4}(V_{A}/L)$,
the fastest growing mode will be amplified by a factor of $\sim\exp(S^{1/4})$
before being convected out of the layer. If the initial disturbance
is sufficiently small, it may not grow to a perceptible size, and
the current sheet will appear to be stable. For this reason, we do
not expect a clear-cut critical value of $S$ for the instability.
Simulations typically give a critical value $S_{c}\sim10^{4}$,\cite{Biskamp1986,BhattacharjeeHYR2009,CassakSD2009,HuangB2010,LoureiroSSU2012}
although clean simulations that remain stable up to $S\sim10^{5}$
have been reported.\cite{NgR2011} {There are also
indications that the critical value $S_{c}$ may depend on plasma
$\beta$.\cite{NiZHLM2012} }Because the {linear}
amplification factor increases rapidly as $S$ becomes large (e.g.,
$\exp(S^{1/4})\sim10^{8},10^{14},10^{24}$ at $S=10^{5},10^{6},10^{7}$,
respectively), the Sweet-Parker current sheet becomes very fragile
at high $S$, and plasmoid formation, which is the consequence of
a true physical instability, is unavoidable. {For
instance, while noise levels significantly lower than $10^{-8}$ of
the background field may be sufficient to obtain a stable Sweet-Parker
current sheet at $S=10^{5}$, it requires a noise level significantly
lower than $10^{-24}$ to achieve a similar level of stability at
$S=10^{7}$. An even more stringent upper bound on the noise level
is obtained at higher values of $S$. We should point out that the
linear amplification factor is introduced here to estimate the requirement
on noise in order to keep the Sweet-Parker current sheet stable. It
may not be representative of the real amplification, because the initial
exponential growth is expected to slow down when nonlinear effects
become important, which typically occurs once the the plasmoid size
becomes comparable to the current sheet width.\cite{Rutherford1973} }

\section{Effects of Plasmoid Instability in Collisional Magnetic Reconnection
\label{sec:Nonlinear}}

\begin{figure*}
\begin{centering}
\includegraphics[scale=0.9]{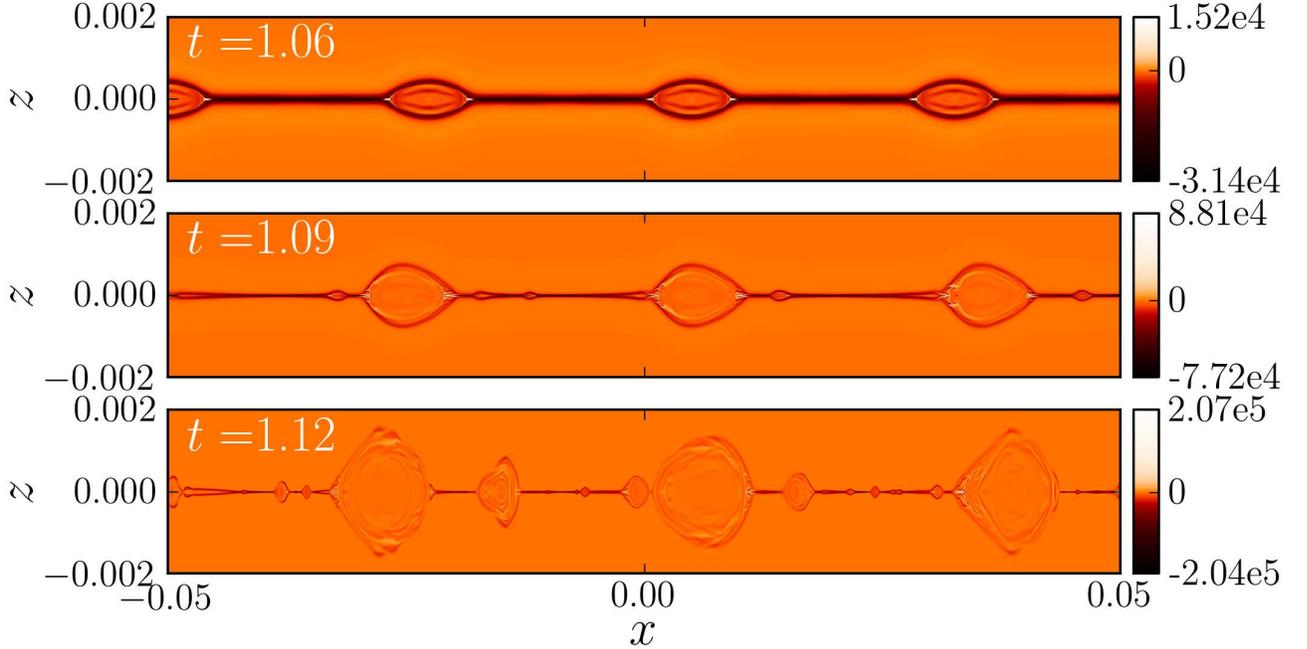}
\par\end{centering}

\caption{(Color online) Out-of-plane current density at different times shows
fractal-like cascade to smaller scales via the plasmoid instability.
First, the Sweet-Parker current sheet breaks up to form a chain of
plamoids connected by secondary current sheets (top panel). Secondary
current sheets are Sweet-Parker like, become unstable again, and generate
the next batch of plasmoids (middle panel). This cascade leads to
a hierarchy of plasmoids of various sizes (bottom panel). These snapshots
present a small portion of the whole simulation box from a $S=10^{7}$
simulation. The reader is referred to Fig. 4 of Ref. \cite{HuangB2010}
for an illustration of the whole system. \label{fig:Fractal-like-cascade}}
\end{figure*}

After onset of the plasmoid instability, the reconnection layer changes
to a chain of plasmoids connected by secondary current sheets that,
in turn, may become unstable again. This process of cascading to smaller
scales is reminiscent of fractals.\cite{ShibataT2001} {If
the large-scale configuration evolves slowly}, eventually the reconnection
layer will tend to a statistical steady state characterized by a hierarchical
structure of plasmoids (see Fig. \ref{fig:Fractal-like-cascade} for
snapshots of the cascade). {Two-dimensional} numerical
simulations show that reconnection rate becomes nearly independent
of $S$ in this regime, with a value $\sim0.01V_{A}B$.\cite{BhattacharjeeHYR2009,HuangB2010,LoureiroSSU2012}
The number of plasmoids $n_{p}$, the widths $\delta$ and lengths
$l$ of secondary current sheets follow scaling relations $n_{p}\propto S$,
$\delta\propto1/S$, and $l\propto1/S$.\cite{HuangB2010} These scaling
relations may be understood by assuming that all secondary current
sheets are close to marginal stability. The rationale behind the assumption
is as follows. Firstly, we note that cascade to smaller scales will
stop if the local Lundquist number $lV_{A}/\eta$ of a secondary current
sheet is smaller than $S_{c}$. Secondly, secondary current sheets
typically get stretched and become longer over time due to gradient
in the outflow, which on average increases from zero near the center
to $\sim V_{A}$ near the ends of the reconnection layer. Thirdly,
when the current sheet length $l$ becomes sufficiently long such
that $lV_{A}/\eta>S_{c}$, the current sheet becomes unstable, new
plasmoids are generated, and the fragmented current sheets become
short again. Consequently, we expect the local Lundquist number $lV_{A}/\eta$
to stay close to $S_{c}$, i.e. $l\sim\eta S_{c}/V_{A}\sim LS_{c}/S$.
The corresponding current sheet width $\delta\sim l/\sqrt{S_{c}}\sim LS_{c}^{1/2}/S$,
and the number of plasmoids is estimated as $n_{p}\sim L/l\sim S/S_{c}$.
Finally, reconnection rate can be estimated by $\eta J\sim\eta B/\delta\sim BV_{A}/\sqrt{S_{c}}\sim10^{-2}V_{A}B$,
which is independent of $S$. These scaling relations are consistent
with the simulation results.

\section{Statistical Distribution of Plasmoids\label{sec:Statistical-Distribution}}

Statistical descriptions of plasmoids have drawn considerable interest
in recent years, \cite{FermoDS2010,UzdenskyLS2010,FermoDSH2011,LoureiroSSU2012}
partly due to the possible link between plasmoids and energetic particles.\cite{DrakeSCS2006,ChenBPYBIMDLKVFG2008}
The fractal-like fragmentation of current sheets suggests self-similarity
across different scales,\cite{ShibataT2001} which often gives rise
to power laws.\cite{Schroeder1990} A recent heuristic argument by
Uzdensky \emph{et al.} suggests that if we consider the statistical
distribution of the plasmoids in terms of their magnetic fluxes $\psi$,
the distribution function $f(\psi)$ follows a power law $f(\psi)\sim\psi^{-2}$.\cite{UzdenskyLS2010}
This result can be formally derived by adopting a model of plasmoid
kinetics (similar to that given in Ref. \cite{FermoDS2010}) and obtaining
steady-state solutions of the plasmoid distribution.\cite{HuangB2012} 

The governing kinetic equation for the time evolution of $f(\psi)$
is written as 

\begin{equation}
\frac{\partial f}{\partial t}+\alpha\frac{\partial f}{\partial\psi}=\zeta\delta(\psi)-\frac{fN}{\tau_{A}}-\frac{f}{\tau_{A}},\label{eq:governing_eq}
\end{equation}
where $N(\psi)\equiv\int_{\psi}^{\infty}f(\psi')d\psi'$ is the cumulative
distribution function, i.e. the number of plasmoids with fluxes larger
than $\psi$. Several idealized assumptions have been made in writing
Eq. (\ref{eq:governing_eq}). Firstly, the flux of a plasmoid grows
due to reconnection in adjacent secondary current sheets. Following
the assumption that all secondary current sheets are close to marginal
stability, the flux of a plasmoid grows approximately at a constant
rate $\alpha\sim BV_{A}/\sqrt{S_{c}}$. This gives the plasmoid growth
term $\alpha\partial f/\partial\psi$ on the left hand side. Secondly,
new plasmoids are created when a secondary current sheet becomes longer
than the critical length for marginal stability. We assume that when
new plasmoids are created, they contain zero flux; this is represented
by the source term $\zeta\delta(\psi$), where $\delta(\psi)$ is
the Dirac $\delta$-function, and $\zeta$ is the magnitude of the
source. This source term sets the boundary condition for $f(\psi)$
at $\psi=0$. Thirdly, plasmoids disappear due to coalescence with
larger plasmoids, which is assumed to be instantaneous. Assuming the
characteristic relative velocity between plasmoids is of the order
of $V_{A}$, the time scale of a plasmoid with flux $\psi$ to encounter
a larger plasmoid is estimated as $\sim L/N(\psi)V_{A}\sim\tau_{A}/N(\psi)$.
This gives the coalescence loss term $-fN/\tau_{A}$. Note that when
two plasmoids coalesce, the flux of the merged plasmoid is equal to
the larger of the two original fluxes.\cite{FermoDS2010} Therefore,
coalescence does not affect the value of $f$ at the larger of the
two fluxes. Lastly, plasmoids are advected out from the reconnection
layer with speeds $\sim V_{A}$ on a characteristic time scale $\tau_{A}$.
This is represented by the advection loss term $-f/\tau_{A}$.

Exact steady-state solutions of Eq. (\ref{eq:governing_eq}) can be
found analytically.\cite{HuangB2012} However, for the discussion
here it is instructive to consider approximate solutions in different
regimes. At large $\psi$ when $N\ll1$, the steady-state equation
reduces to $\alpha\partial f/\partial\psi\simeq-f/\tau_{A}$. In this
regime $f\sim\exp(-\psi/\alpha\tau_{A})$. On the other hand, when
$N\gg1$, the advection loss term is negligible, and we have $\alpha\partial f/\partial\psi\simeq-fN/\tau_{A}$.
In this regime, $N\simeq2\alpha\tau_{A}\psi^{-1}$ and $f=-\partial N/\partial\psi\simeq2\alpha\tau_{A}\psi^{-2}$
is the solution. As such, the steady state solution admits both an
exponential tail and a $f\sim\psi^{-2}$ power-law regime. The dominant
loss mechanism in the former regime is advection, while it is coalescence
in the latter. In other words, the plasmoids in the power-law regime
must be deep in the hierarchy, whereas large plasmoids follow a distribution
that falls off exponentially. Transition from the power-law regime
to the exponential tail occurs when $N\simeq1$, i.e. at $\psi\simeq2\alpha\tau_{A}$.
The distribution function $f(\psi)$ also deviates from the $\psi^{-2}$
power law in the small $\psi$ limit, otherwise the cumulative distribution
function $N(\psi)$ will diverge as $\psi\to0$. Because $N(\psi)\to n_{p}\sim S/S_{c}$
as $\psi\to0$, the transition occurs when $2\alpha\tau_{A}\psi^{-1}\simeq n_{p}$,
i.e. when $\psi\simeq2\alpha\tau_{A}/n_{p}$. Therefore, the power
law holds in the range $2\alpha\tau_{A}/n_{p}\ll\psi\ll2\alpha\tau_{A}$,
which becomes more extended for higher $S$.

\begin{figure}[t]
\begin{centering}
\includegraphics[scale=0.9]{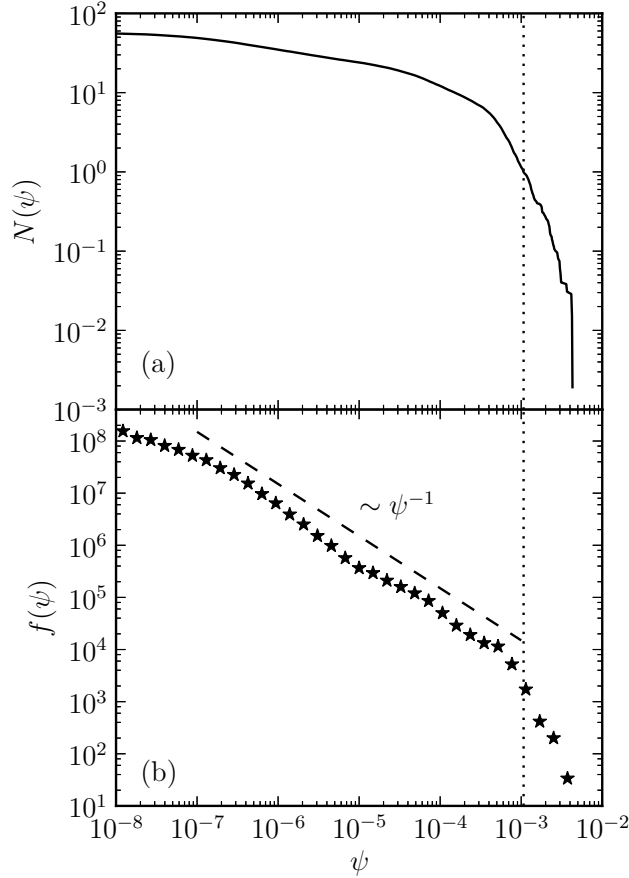}
\par\end{centering}

\caption{(a) Cumulative distribution function $N(\psi)$ and (b) distribution
function $f(\psi)$ of plasmoids from a $S=10^{7}$ simulation. The
vertical dotted line denotes where $N(\psi)=1$. \label{fig:Plasmoid-distributions-from-DNS}}
\end{figure}

This prediction of $f(\psi)\sim\psi^{-2}$ power-law distribution
can be tested with numerical simulations. Figure \ref{fig:Plasmoid-distributions-from-DNS}
shows the cumulative distribution function $N(\psi)$ (panel (a))
and the distribution function $f(\psi)$ (panel (b)) from a $S=10^{7}$
simulation reported in Ref. \cite{HuangB2012}. The dataset was constituted
of 30507 plasmoids collected from an ensemble of 521 snapshots during
the quasi-steady phase with a cadence of 100 snapshots per $\tau_{A}$;
an ensemble average was carried out for better statistics. The distribution
function $f(\psi)$ exhibits an extended power-law regime; however,
the power law is close to $f(\psi)\sim\psi^{-1}$ instead of $f(\psi)\sim\psi^{-2}$.
The vertical dotted line denotes where $N(\psi)$ crosses $N=1$,
indicating the switch of the dominant loss mechanism from coalescence
to advection. {From the above analysis, this switch
of the dominant loss mechanism is responsible for the transition from
a power-law distribution to an exponential falloff. And indeed, where
$N(\psi)$ crosses $N=1$ approximately coincides with where the distribution
function $f(\psi)$ starts to deviate from $f(\psi)\sim\psi^{-1}$
to a more rapid falloff.} Incidentally, this rapidly falling tail
was where Loureiro \emph{et al. }attempted to fit with the $f(\psi)\sim\psi^{-2}$
power law.\cite{LoureiroSSU2012} At smaller $\psi$, their reported
distribution also appears to be consistent with $f(\psi)\sim\psi^{-1}$.
{Because our simulation only lasted a few $\tau_{A}$,
the statistics in the large-$\psi$ regime is sufficiently uncertain
that it is difficult to make a clear distinction between a $\psi^{-2}$
and an exponential falloff. Observationally, the distribution of the
flux transfer events (FTEs) in the magnetopause from Cluster data,
collected over a period of two years, appears to be consistent with
an exponential tail.\cite{FermoDSH2011} }

\begin{figure}
\begin{centering}
\includegraphics[bb=0bp 0bp 245bp 210bp,clip,scale=0.9]{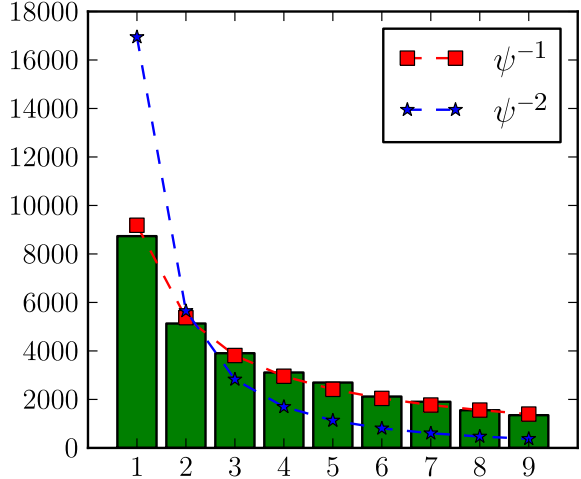}
\par\end{centering}

\caption{(Color online) The distribution of the leading digit of flux shows
good agreement with Benford's law (for $f(\psi)\sim\psi^{-1}$), but
deviates significantly from the prediction based on $f(\psi)\sim\psi^{-2}$.
\label{fig:Benford}}
\end{figure}

An alternative way to make a distinction between $\psi^{-1}$ and
$\psi^{-2}$ distributions is to examine the distribution of the leading
digit $d$ of the flux. (For example, if $\psi=2.35\times10^{-5}$,
then $d=2$.) Although the leading digit surely depends on the unit
we use, the distribution of the leading digit will remain the same
if the underlying distribution is a power law. For the $f(\psi)\sim\psi^{-1}$
distribution, the probability $P(d)$ of the leading digit $d$ follows
Benford's law $P(d)=\log_{10}(1+1/d)$.\cite{Benford1938} On the
other hand, the probability will be $P(d)=10/9d(d+1)$ if the distribution
is $f(\psi)\sim\psi^{-2}$. Figure \ref{fig:Benford} shows the distribution
of the leading digit, which is in good agreement with Benford's law,
but deviates significantly from the prediction corresponding to the
distribution $f(\psi)\sim\psi^{-2}$. Because all plasmoids in the
dataset, not just those in the power-law regime, are used in this
analysis, the good agreement with Benford's law reflects the fact
that the majority of the plasmoids are in the $f(\psi)\sim\psi^{-1}$
power-law regime. 

\begin{figure}
\begin{centering}
\includegraphics[scale=0.8]{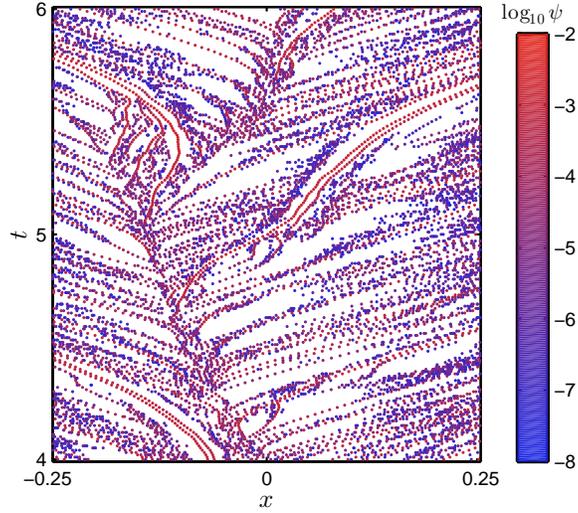}
\par\end{centering}

\caption{(Color online) Stack plot of plasmoid positions along the outflow
direction $x$ during the period $t=4$ to $t=6$. Each dot represents
a plasmoid, color coded according to its flux in logarithmic scale.
\label{fig:stack}}
\end{figure}

As we have discussed, the dominant balance in Eq. (\ref{eq:governing_eq})
leading to the $f(\psi)\sim\psi^{-2}$ solution is between the plasmoid
growth term and the coalescence loss term, i.e. $\alpha\partial f/\partial\psi\simeq-fN/\tau_{A}$.
A key assumption underlying the loss term $-fN/\tau_{A}$ is that
the relative speeds of a plasmoid with respect to neighboring plasmoids
larger than itself are of the order of $V_{A}$ and are uncorrelated
to the flux of the plasmoid. However, that was found not to be the
case when we examined the numerical data. Instead, the relative speeds
between large plasmoids tend to be lower.\cite{HuangB2012} Figure
\ref{fig:stack} shows the stack plot of plasmoid positions along
the outflow direction $x$ over a period of $2\tau_{A}$, where each
dot represents a plasmoid, color coded according to its flux in logarithmic
scale. The stack plot has a tree-like structure, with branches that
represent the trajectories of larger plasmoids shown on the red side
of color scale. Those dots (or ``leaves'') on the blue side of the
color scale are mostly smaller plasmoids that do not show perceivable
trajectories. We can clearly see that many branches are nearly parallel
to their neighboring branches, indicating low relative speeds between
large neighboring plasmoids. 

This interesting phenomenon may be understood as follows. Roughly
speaking, the flux of a plasmoid is proportional to its age because
all plasmoids approximately grow at the same rate $\alpha$. Consequently,
a plasmoid can become large only if it has not encountered plasmoids
larger than itself for an extended period of time. Those plasmoids
moving rapidly relative to their neighbors will encounter larger plasmoids
and disappear easily, whereas those with small relative speeds are
more likely to survive for a long time and become large. To incorporate
this important effect, we have to consider a distribution function
not only in flux, but also in velocity as well. Let $F(\psi,v)$ be
the new distribution function, where $v$ is interpreted as the plasmoid
velocity relative to the mean flow. We propose the following kinetic
model for $F(\psi,v)$: 
\begin{equation}
\partial_{t}F+\alpha\frac{\partial F}{\partial\psi}=\zeta\delta(\psi)h(v)-\frac{FH}{\tau_{A}}-\frac{F}{\tau_{A}},\label{eq:modified}
\end{equation}
where the function $H$ is defined as 
\begin{equation}
H(\psi,v)=\int_{\psi}^{\infty}d\psi^{'}\int_{-\infty}^{\infty}dv'\frac{\left|v-v'\right|}{V_{A}}F(\psi',v'),\label{eq:collision}
\end{equation}
and $h(v)$ is an arbitrary distribution function in velocity space
when new plasmoids are generated. The distribution function $f(\psi)$
can be obtained by integrating $F(\psi,v)$ over velocity space. Eq.
(\ref{eq:modified}) differs from Eq. (\ref{eq:governing_eq}) in
the coalescence loss term, where the relative speed $\left|v-v'\right|$
between two plasmoids is taken into account in the integral operator
of Eq. (\ref{eq:collision}). If we replace $\left|v-v'\right|$ in
Eq. (\ref{eq:collision}) by $V_{A}$, then Eq. (\ref{eq:modified})
reduces to Eq. (\ref{eq:governing_eq}) after integrating over velocity
space. By numerically solving for steady-state solutions of Eq. (\ref{eq:modified}),
we find that the distribution in the power-law regime is close to
$f(\psi)\sim\psi^{-1}$, consistent with DNS results (see Figure 4
of Ref. \cite{HuangB2012}). This conclusion does not appear to be
sensitive to the specific form of $h(v)$, as long as $h(v)$ covers
a broad range of $v$ (typically of the order of $V_{A}$). 

\begin{figure}
\begin{centering}
\includegraphics[scale=0.8]{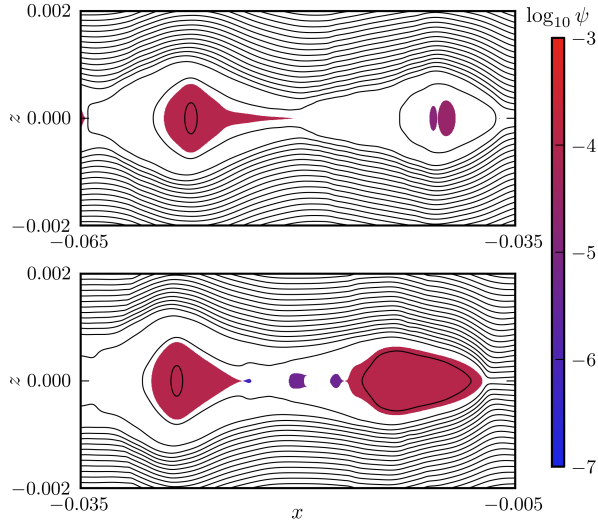}
\par\end{centering}

\caption{(Color online) Top: On the right hand side of the panel is an example
of two partially merged plasmoids, both small in size. Bottom: When
the merging is completed, they become a plasmoid much larger in extent.
Also there are three new plasmoids generated in the middle. Contour
lines represent magnetic field lines, and plasmoids are color coded
with fluxes in logarithmic scale. Note that the two panels cover different
ranges in the $x$ direction, because the whole structure is moving.
\label{fig:partial-merge}}

\end{figure}

We caution the reader that although the heuristic argument in Sec.
\ref{sec:Nonlinear} and the kinetic model in this Section appear
to account for the numerically observed scaling laws and plasmoid
distributions, they are far from complete and satisfactory descriptions
of the complex dynamics of a plasmoid-dominated reconnection layer.
In particularly, a key building block of the present theory is the
assumption that plasmoids are connected by marginally stable current
sheets. That is clearly an oversimplification and does not appear
to be always (and quite often not) the case upon close examination
of simulation data;\cite{HuangB2010} therefore, further exploration
is warranted.\cite{Baty2012} Some other potentially important effects
are also omitted. {For instance, plasmoids are treated
as point ``particles'' and coalescence between them is assumed to
occur instantaneously, whereas in reality larger plasmoids take longer
to merge, and there can be bouncing (or sloshing) between them.\cite{KnollC2006,KarimabadiDRDC2011}
Furthermore, the velocity $v$ relative to the mean flow is assumed
to remain constant throughout the lifetime of a plasmoid, whereas
in reality some variation is expected due to the complex dynamics
between plasmoids. }

{The fact that plasmoids can be in a partially merged
state can have significant effects on the statistical distribution,\cite{LoureiroSSU2012}
and this issue is further complicated by that there are different
ways of identifying the extent of a plasmoid among researchers. In
our diagnostics, plasmoids are extrema (O-points) of the flux function,
which we solve as a primary variable in the simulation code. The extent
of a plasmoid is determined by expanding the level set of the flux
function from the extremum until it reaches a saddle point (X-point).
Note that our convention differs from that adopted by Fermo }{\emph{et
al}}{. (see Figure 3 of Ref. \cite{FermoDSH2011}).
Our method treats all plasmoids on equal footing, whereas that of
Fermo }{\emph{et al}}{. makes a
distinction between the dominant and the lesser plasmoids for partially
merged plasmoids. Although our method is mathematically unambiguous,
it has a consequence that when two plasmoids are in the process of
merging, both shrink in size until the merging is completed and a
large plasmoid appears suddenly (Figure \ref{fig:partial-merge}),
whereas in the convention of Fermo}{\emph{ et al.}}{{}
the lesser plasmoid shrinks in size and the dominant one keeps growing.
This is a subtle point that merits further consideration if we apply
the plasmoid distribution functions to the problem of particle energization.}

{}

{}

\section{Roles of Plasmoid Instability in Onset of Collisionless Reconnection
\label{sec:Roles-of-Plasmoid}}

\begin{figure}
\begin{centering}
\includegraphics{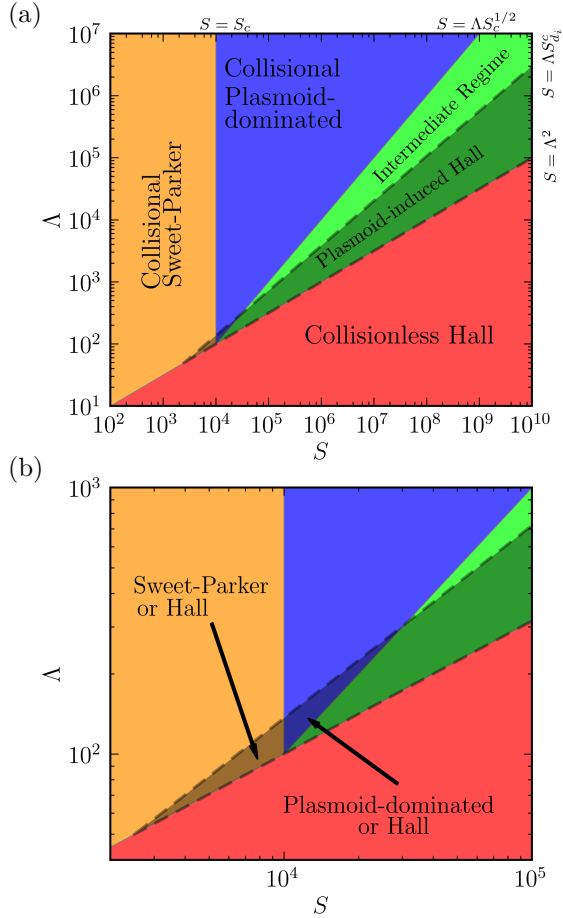}
\par\end{centering}

\caption{(Color online) (a) Phase diagram of magnetic reconnection that shows
the five phases. (b) Expanded view about $S=10^{4}$ and $\Lambda=10^{2}$
shows two history-dependent regimes.\label{fig:Phase-diagram}}
\end{figure}

{Thus far, our discussion assumes that resistive MHD
remains valid down to the smallest scales. This assumption is clearly
questionable when current sheet widths reach kinetic scales, when
two-fluid (Hall) and kinetic effects become important.} Conventional
wisdom had it that the onset of fast reconnection occurs when the
Sweet-Parker width $\delta_{SP}$ is smaller than the ion inertial
length $d_{i}$ (which should be replaced by{{} the
ion Larmor radius at the sound speed, $\rho_{s}$, if there is a strong
guide field).\cite{Aydemir1992,MaB1996a,DorelliB2003,Bhattacharjee2004,CassakSD2005,CassakDS2007}
Because secondary current sheets can be much thinner than the Sweet-Parker
width, the implication is that collisionless reconnection may set
in even when the conventional criterion for onset is not met. This
has been confirmed by several recent studies using fully kinetic \cite{DaughtonRAKYB2009}
and Hall MHD \cite{ShepherdC2010,HuangBS2011} models. Along with
these studies, a practice has emerged of using phase diagrams to describe
various possible ``phases'' of reconnection in the parameter space
of $S$ and $\Lambda\equiv L/d_{i}$.\cite{HuangBS2011,JiD2011,DaughtonV2012,CassakD2013}}
{Figure \ref{fig:Phase-diagram} shows our current
rendition of the phase diagram, which is divided into seven regimes
(including two history-dependent regimes), and five ``phases'',
that will be detailed as follows. }

{The collisionless Hall phase is realized when the
conventional criterion for onset is satisfied, i.e. $\delta_{SP}<d_{i}$.
In the parameter space of $S$ and $\Lambda$, that translates to
$S>\Lambda^{2}$. This is the regime where reconnection dominated
by collisionless effects will surely occur. In the region $S<\Lambda^{2}$,
collisional Sweet-Parker reconnection can be realized if the current
sheet is stable, i.e. when $S<S_{c}$. However, that is not always
the case. An important phenomenon discovered by Cassak }{\emph{et
al. }}{is that reconnection is not uniquely determined
by the parameters $S$ and $\Lambda$ alone; in a so called ``bistable''
region of the parameter space, both Sweet-Parker and Hall reconnection
are possible, depending on the history in parameter space.\cite{CassakSD2005,CassakDS2007,CassakSD2010,SullivanBH2010}
A useful parameter to discuss this hysteresis phenomenon }is the Lundquist
number based on $d_{i}$, defined as $S_{d_{i}}\equiv V_{A}d_{i}/\eta=S/\Lambda$,
which characterizes the effect of resistivity on the $d_{i}$ scale.
Disregarding the plasmoid instability, the condition for bistability
may be expressed as $\Lambda>S_{d_{i}}>S_{d_{i}}^{c},$ or equivalently
\begin{equation}
\Lambda S_{d_{i}}^{c}<S<\Lambda^{2},\label{eq:bistable}
\end{equation}
where $S_{d_{i}}^{c}$ is a critical value of $S_{d_{i}}$. The value
of $S_{d_{i}}^{c}$ is found to be $\sim O(10^{2})$;\cite{CassakSD2005,HuangBS2011}
however, numerical evidence indicates that $S_{d_{i}}^{c}$ may increase
with increasing $\Lambda$,\cite{HuangBS2011} although the precise
scaling is not known. The parameter space defined by Eq. (\ref{eq:bistable})
is represented by the shaded region enclosed by dashed lines in Fig.
{\ref{fig:Phase-diagram}, where we have assumed
an arbitrary scaling just to show qualitatively the dependence of
$S_{d_{i}}^{c}$ on $\Lambda$. The regime in the overlapping region
defined by }Eq. (\ref{eq:bistable}){{} and $S<S_{c}$
is the true ``bistable'' regime, where either collisionless Hall
or collisional Sweet-Parker reconnection can be realized, depending
on the history (see panel (b) of }Fig. {\ref{fig:Phase-diagram}
for a zoom-in view). }

When $S>S_{c}$, the Sweet-Parker current sheet is unstable. However,
reconnection may remain collisional provided that secondary current
sheet widths stay above the $d_{i}$ scale. Using the scaling estimate
$\delta\sim LS_{c}^{1/2}/S$, the criterion is 
\begin{equation}
S_{c}<S<\Lambda S_{c}^{1/2},\label{eq:plasmoid-diminated}
\end{equation}
which defines the collisional plasmoid-dominated phase. Note that
the there is an overlapping region defined by Eq. (\ref{eq:bistable})
and Eq. (\ref{eq:plasmoid-diminated}) together ({again
see panel (b) of }Fig. {\ref{fig:Phase-diagram}
for a zoom-in view}). That suggests there may exist a history-dependent
regime where both collisionless Hall and collisional plasmoid-dominated
reconnection can be realized. Although this regime has not been demonstrated
in simulation, it remains an interesting possibility.

The region defined by $\Lambda S_{c}^{1/2}<S<\Lambda^{2}$ is where
secondary current sheets can reach the $d_{i}$ scale. {Where
this region overlaps the ``bistable'' region defined by Eq. (\ref{eq:bistable})
is the ``plasmoid-induced Hall'' phase. Although this regime is
supposed to be bistable according to Eq. (\ref{eq:bistable}), Sweet-Parker
reconnection ceases to exist due to the plasmoid instability, and
collisionless Hall reconnection is the only possibility. The only
difference between the ``plasmoid-induced Hall'' regime and the
``collisionless Hall'' regime is the process of reaching collisionless
Hall reconnection. In the former, onset of Hall reconnection is proceeded
by cascading to smaller scales induced by plasmoids, whereas in the
latter the primary current sheet will reach the kinetic scales and
immediately trigger the onset of Hall reconnection.}

The region defined by $\Lambda S_{c}^{1/2}<S<\Lambda S_{d_{i}}^{c}$
is the ``intermediate regime'' {where the system
alternates between collisional and collisionless reconnection}. On
the one hand, secondary current sheets can reach the $d_{i}$ scale
and trigger onset of Hall reconnection. On the other hand, after onset
of Hall reconnection the system cannot settle down to a localized
Hall reconnection geometry because it is physically unrealizable.
This regime has been realized in a large scale Hall MHD simulation,
where the current sheet becomes extended again after onset of Hall
reconnection. That leads to formation of new plasmoids{{}
and another onset of Hall reconnection.}\cite{HuangBS2011}{{} }

{From what we have learned so far, }among these phases,
the collisional plasmoid-dominated regime gives the slowest reconnection
rate, which is approximately $0.01V_{A}B$. The collisional Sweet-Parker
regime yields faster reconnection, because resistivity is high. Collisionless
Hall reconnection rate typically attains a maximum value $\sim0.1V_{A}B$,
whereas in the intermediate regime the reconnection rate can oscillate
between $0.01V_{A}B$ and $0.1V_{A}B$. 

There have been different renditions of phase diagrams proposed in
recent literature. {\cite{HuangBS2011,JiD2011,DaughtonV2012,CassakD2013}
All of them share a great similarity, with some minor differences.
Because the parameter space has not been systematically explored,
these phase diagrams are necessarily speculative to some extent. Nonetheless,
they can serve as a good frame of reference to explore large scale
magnetic reconnection, either in choosing simulation parameters \cite{HuangBS2011}
or in planning for future experiments.\cite{JiD2011} One should bear
in mind that the border lines between different phases are not ironclad,
because there is no clear-cut value of $S_{c}$. Furthermore, secondary
current sheet widths can deviate from the estimated value }$\delta\sim LS_{c}^{1/2}/S$.
Finally, both $S_{c}$ and $\delta$ can be affected by Hall effects.\cite{BaalrudBHG2011}
More caveats in applying the phase diagrams are discussed in Ref.
\cite{CassakD2013}, and a compilation of parameters for various astrophysical,
space, and laboratory plasmas can be found in Ref. \cite{JiD2011}.

\section{Discussion and Future Challenges\label{sec:Discussion-and-Future}}

Although significant advances have been made in recent years on large
scale, high-Lundquist-number reconnection, where plasmoids play important
roles, many questions remain open. The results presented here are
based on 2D studies. To what extent do these 2D results carry over
to 3D geometry, where oblique tearing modes have been shown to play
an important role?\cite{DaughtonRKYABB2011,BaalrudBH2012} The relationship
between plasmoid-dominated reconnection and turbulent reconnection\cite{LazarianV1999,KowalLVO2009,EyinkLV2011,LazarianEV2012}
is another issue that needs further investigation. What are their
similarities and differences? Will the interaction between overlapping
oblique modes in 3D lead to self-generated turbulence and further
blur the line between the two?

How do global conditions affect the plasmoid instability and magnetic
reconnection in general also needs further assessment. One of the
most important results from recent 2D studies is that reconnection
rate cannot be significantly slower than $10^{-2}V_{A}B$, therefore
is always fast {(although this has only been tested
up to $S\sim10^{7}$)}. This conclusion poses new challenges to theories
of solar flares that assume the existence of slow reconnection.\cite{Uzdensky2007,Uzdensky2007a,CassakMS2008,CassakS2012,CassakD2013}
However, these studies were done in simple 2D configurations, and
it is not clear whether the results can be directly applied to solar
corona, where the line-tying effect is thought to play an important
role.\cite{GibonsS1981,Hood1986,Hood1992,LongcopeS1994} Line-tying
is known to have stabilizing effect on tearing instability \cite{DelzannoF2008,HuangZ2009}
and smoothing effect on current sheets.\cite{EvstatievDF2006,HuangZS2006,HuangBZ2010}
Both effects could significantly affect the plasmoid instability.
{Although line-tying has been employed in the lower
boundary of some recent 2D simulations of the plasmoid instability
in coronal current sheets, \cite{BartaBKS2011,ShenLM2011,MeiSWLMR2012}
these studies are limited to anti-parallel reconnection, where line-tying
stabilization is not effective. Line-tying stabilization of the tearing
instability is more effective for component reconnection, with the
guide field line-tied to the boundary. Therefore, to study the effects
of line-tying on the plasmoid instability, 3D simulations will be
needed.}

Even in 2D systems, the current understanding is far from complete.{{}
The parameter space needs to be systematically explored with higher
Lundquist number and larger system size than that have been achieved,
to test the proposed phase diagrams. The kinetic model of plasmoid
distribution can be further improved by including other coarse-graining
variables in addition to $\psi$ and $v$, and the distribution of
plasmoids in collisionless regime needs to be further studied. }
\begin{acknowledgments}
This work was supported {by the Department of Energy,
Grant No. DE-FG02-07ER46372, under the auspice of the Center for Integrated
Computation and Analysis of Reconnection and Turbulence (CICART),
the National Science Foundation, Grant No. PHY-0215581 (PFC: Center
for Magnetic Self-Organization in Laboratory and Astrophysical Plasmas),
NASA Grant Nos. NNX09AJ86G and NNX10AC04G, and NSF Grant Nos. ATM-0802727,
ATM-090315 and AGS-0962698. Computations were performed on Oak Ridge
Leadership Computing Facility through an INCITE award, and National
Energy Research Scientific Computing Center.}
\end{acknowledgments}
\bibliographystyle{apsrev4-1}

\end{document}